\documentclass[english,12pt,a4]{article}
\usepackage[T1]{fontenc}
\usepackage[latin1]{inputenc}
\usepackage{a4}
\usepackage{amsmath}
\usepackage{amssymb,amsmath}
\usepackage{epsfig}
\usepackage{graphics,graphicx}

\makeatletter

\providecommand{\LyX}{L\kern-.1667em\lower.25em\hbox{Y}\kern-.125emX\@}

 \newcommand{\lyxaddress}[1]{
   \par {\raggedright #1 
   \vspace{1.4em}
   \noindent\par}
 }

\makeatother
\begin{document}

\title{\textbf{\Large On axially symmetrical solitons in Abelian-Higgs models }}

\author{{\normalsize C G Doudoulakis}}

\date{{}}

\maketitle

\lyxaddress{Department of Physics and Institute of Plasma Physics, University of Crete; Heraklion, Crete, Greece}

\begin{abstract}
A numerical search for bosonic superconducting static vortex rings in a $U(1)_{A}\times U(1)_{W}$ model is presented.
The fate of these rings without current, is to shrink due to their tension until extinction.
The superconductivity of the loop does not seem to prevent shrinking.
Current quenching takes place before stabilization.

\end{abstract}

\maketitle

\section {{\normalsize Introduction}}
In a series of papers \cite{c1, c2} classically stable, metastable quasi-topological domain walls and strings in simple topologically trivial models,
as well as in the two-higgs Standard Model (2HSM) were studied. They are local minima of the energy functional and can quantum
mechanically tunnel to the vacuum, not being protected by an absolutely conserved quantum number. In \cite{c3} a search for
spherically symmetric particle like solitons in the 2HSM with a simplified higgs potential was performed without success.
Although the existence of spherically symmetric particle-like solitons in the 2HSM has not been ruled out, we shall here look, instead,
for axially symmetric solutions in a similar system.

\begin{figure}[t]
\centering
\includegraphics[scale=0.52]{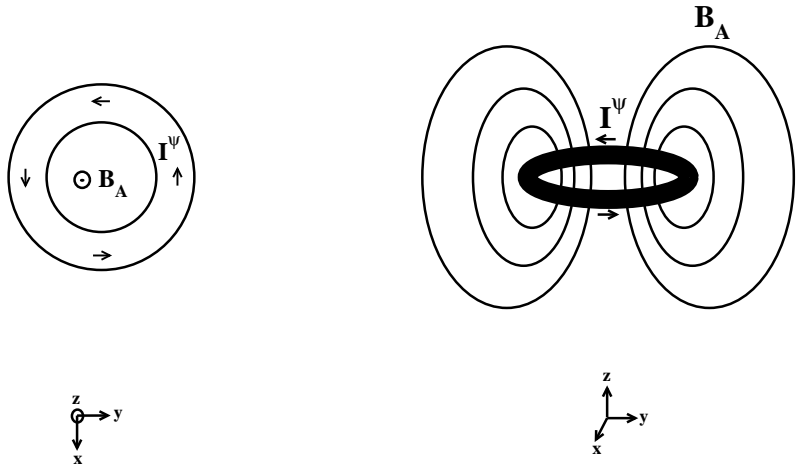}
\caption{\small A $x-y$ profile of the superconducting ring (left) as well as a $y-z$ profile (right)
where one can view how the mechanism we propose, against shrinking, works.\label{mechh} }
\end{figure}

Consider a model with superconducting strings \cite{c2, c4}. Take a piece of such string, close it to form a donut-shaped loop and let current in it.
A magnetic field due to the supercurrent will be passing through the hole of the donut (fig.\ref{mechh}). The energy of  the loop has, a term
proportional to the length of the string and will tend to shrink the radius of the donut to zero and the ring to extinction. However,
another force opposes this tendency. Namely, as the loop shrinks, the magnetic field lines are squeezed in the hole, since, due to 
the Meissner effect, they cannot leave the loop. They are trapped inside the hole of the donut, oppose further shrinking and might 
even stabilize the string.

This, as well as other arguments \cite{c7} are inspiring but not conclusive. The magnetic field will not be trapped inside the loop if the penetration depth
of the superconductor is larger than the thickness of the ring. Also, once the magnetic field gets strong  it can destroy
superconductivity and penetrate \cite{c8}. Further, there is a maximum current a superconductor can support (current quenching).
This sets a limit on the magnetic field one can have through the loop, and this may not be enough to stabilize it.
Thus, the above approach may work at best  in a certain region of the parameter space, depending also on the defect characteristics \cite{c2}.
The purpose of this work is to apply the above straightforward idea to search for string loops \cite{c10}-\cite{c19} in a $U(1)\times U(1)$ gauge model
and to determine the parameter space, if any, for their existence and stability. Another interesting subject is to have a rotating ring. The rotation
is another extra factor which could help the ring to stabilize. This work was done with success both analytically and numerically on \cite{c7b}
where vortons are exhibited. Another recent example of rotating superconducting electroweak strings can be found on \cite{c8a}, while for
a review on electroweak strings, the reader should also check \cite{c8b}. Finally, a work on static classical vortex rings in
$SU(2)$ non-abelian Yang-Mills-Higgs model can be found on \cite{c8c}.

\section {{\normalsize The model}}
The {\bf Lagrangian} describing our system is:
\begin{equation}
\mathcal{L}=-\frac{1}{4}F_{\mu\nu}^{2}-\frac{1}{4}W_{\mu\nu}^{2}+|D_{\mu}\psi|^{2}+|\tilde{D}_{\mu}\phi|^{2}-U(|\phi|,|\psi|)
\end{equation}
where the covariant derivatives are $D_{\mu}\psi \equiv \partial_{\mu}\psi+ieA_{\mu}\psi$, $\tilde{D}_{\mu}\phi \equiv \partial_{\mu}\phi+iqW_{\mu}\phi$,
the strength of the fields are $F_{\mu\nu}=\partial_{\mu}A_{\nu}-\partial_{\nu}A_{\mu}$, $W_{\mu\nu}=\partial_{\mu}W_{\nu}-\partial_{\nu}W_{\mu}$, 
while $e$ and $q$ stand as the relevant $U(1)$ charges.
We choose the potential $U$
\begin{equation}
U(|\phi|,|\psi|)=\frac{g_{1}}{4}\big( |\phi|^{2} -v_{1}^{2}\big)^{2}+\frac{g_{2}}{4}\big( |\psi|^{2}- v_{2}^{2}\big)^{2}+\frac{g_{3}}{2}|\phi|^{2}|\psi|^{2}-\frac{g_{2}}{4}v_{2}^{4}
\end{equation}
The vacuum of this theory is $|\phi|= v_{1} \neq 0$, $|\psi|=0$ and breaks $U(1)_{W} \times U(1)_{A} \rightarrow U(1)_{A}$,
giving non-zero mass to $W$. The photon field stays massless. There, $U(v_{1},0)=0$. 
The vacuum manifold $\mathcal{M}$ in this theory is a circle $S^{1}$ and the first homotopy group of $\mathcal{M}$
is $\pi_{1}(\mathcal{M})=\pi_{1}(S^{1})=\mathbf{Z}$ which signals the existence of strings.
In regions where $|\phi|=0$, the field $|\psi|$ is arranged to be
non-vanishing and $U(1)_{W}\times U(1)_{A}\rightarrow U(1)_{W}$. Thus, $U(1)_{A}\rightarrow \mathbf{1}$ and
one may generate an electric current flowing along regions with vanishing $|\phi|$.
Hence, this theory has superconducting strings \cite{c4}.
The vacuum of the theory leaves unbroken the electromagnetic $U(1)_{A}$.
For $g_{3}v_{1}^{2} > g_{2}v_{2}^{2}$ this vacuum is stable, while $g_{1}v_{1}^{4}>g_{2}v_{2}^{4}$
ensures that it is the global minimum of the potential.
The mass spectrum is 
\begin{equation}
m_{A}=0,\;\;m_{W}=q v_{1},\;\;m_{\phi}^{2}=g_{1}v_{1}^{2},\;\;m_{\psi}^{2}=\frac{1}{2}\big(g_{3}v_{1}^{2}-g_{2}v_{2}^{2}\big)
\end{equation}

\section {{\normalsize The $U(1)_{A}\times U(1)_{W}$ model: Search of superconducting vortex rings}}
Configurations with torus-like shape, representing a piece of a $U(1)_{W}\rightarrow \mathbf{1}$ Nielsen-Olesen string \cite{c9},
closed to form a loop, are of interest in this search. Thus, we will  require $\phi$ to vanish on a circle of radius $a$ (the torus radius)
$\phi (\rho =a, z=0)=0$. At infinity ($\rho\rightarrow \infty, z\rightarrow \infty$), we have the vacuum of the theory. This translates to $|\phi|\rightarrow v_{1}$,
$|\psi| \rightarrow 0$.
The {\bf ansatz} for the fields is:
\begin{eqnarray*}
\phi(\rho,\varphi,z)&=&F(\rho,z)e^{iM\Theta(\rho,z)} \\
\psi(\rho,\varphi,z)&=&P(\rho,z)e^{iN\varphi} \\
\mathbf{A}(\rho,\varphi,z)&=&\frac{A_{\varphi}(\rho,z)}{\rho}\;\hat{\varphi} \\
\mathbf{W}(\rho,\varphi,z)&=&W_{\rho}(\rho,z)\;\hat{\rho}+W_{z}(\rho,z)\;\hat{z} 
\end{eqnarray*}
where $M$, $N$ are the winding numbers of the relevant fields and $\hat{\rho}$, $\hat{\varphi}$, $\hat{z}$ are
the cylindrical unit vectors. We use cylindrical coordinates $(t,\rho , \varphi , z)$,
with space-time metric $g_{\mu\nu}= diag(1,-1,-\rho^{2},-1)$.
We define $\Theta(\rho,z)=\arctan(z/(\rho-a))$.
We work in the $A^{0}=0=W^{0}$ gauge.
Especially for the gauge fields, we suppose the above form based on the following reasonable thoughts:
The $\mathbf{W}$-field is the one related to the formation of the string thus, it exists in the constant-$\varphi$ plane.
This means that in general its non-vanishing components will be $W_{\rho}$ and $W_{z}$.
The $\mathbf{A}$-field is the one produced by the supercurrent flowing inside the toroidal object. The current is 
in the $\hat{\varphi}$ direction thus, we in general expect the non-vanishing component to be $A_{\varphi}$.
Finally, as it concerns the amplitude of all the fields, we expect that it is independent of $\varphi$
due to the axial symmetry of torus.

\begin{figure}
\centering
\includegraphics[scale=0.96]{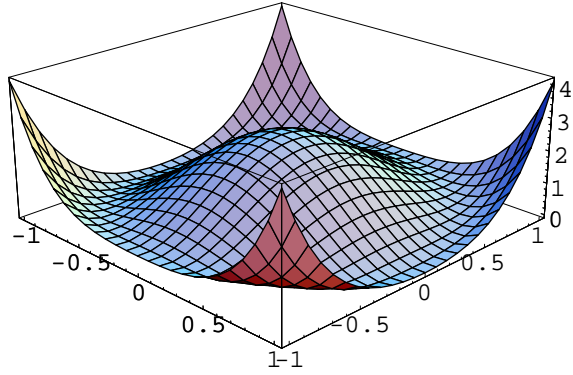}
\caption{\small  A wider view of the ``Mexican hat'' potential (eq.\ref{potu1}) for  $(g_{1},g_{2},u)=(10,8,1)$.\label{mhpos}}
\end{figure}

With the above ansatz, the {\bf energy functional} to be minimized takes the form:
\begin{eqnarray}
\label{funcu1}
E&=&2\pi v_{1} \int_{0}^{\infty}\rho d\rho \int_{-\infty}^{\infty}dz \Bigg[ \frac{1}{2\rho^{2}}\Big( (\partial_{\rho}A_{\varphi})^{2}+(\partial_{z}A_{\varphi})^{2}\Big)+{}
                                                   \nonumber\\
{}&+&(\partial_{\rho}P)^{2}+(\partial_{z}P)^{2}+\frac{P^{2}}{\rho^{2}}(eA_{\varphi}+N)^{2}+{}
                                                          \nonumber\\
{}&+&(\partial_{\rho}F)^{2}+(\partial_{z}F)^{2}+\frac{1}{2} (\partial_{\rho}W_{z}-\partial_{z}W_{\rho})^{2}+{}
                                                  \nonumber\\
{}&+&\Big( (qW_{\rho}+M\partial_{\rho}\Theta)^{2}+(qW_{z}+M\partial_{z}\Theta)^{2}\Big)F^{2}+U(F,P) \Bigg]
\end{eqnarray}
and the potential $U$ (fig.\ref{mhpos}) can be written as follows:
\begin{equation}
\label{potu1}
U(F,P)=\frac{g_{1}}{4}\big( F^{2}-1\big)^{2}+\frac{g_{2}}{4}\big( P^{2}-u^{2}\big)^{2}+\frac{g_{3}}{2}F^{2}P^{2}-\frac{g_{2}}{4}u^{4}
\end{equation}
where  $u\equiv v_{2}/v_{1}$. This is the energy functional we  use for our computations.
The {\bf conditions} to be satisfied by the parameters become:
\begin{equation}
g_{1} > g_{2}u^{4}\;\;\;,\;\;\; g_{3} > g_{2}u^{2}
\end{equation}

The gauge fields $\mathbf{A},\;\mathbf{W}$ have {\bf magnetic fields} of the following form:
\begin{eqnarray*}
\mathbf{\nabla \times A}=\mathbf{B_{A}}= \frac{1}{\rho}\Bigg(\frac{\partial A_{\varphi}}{\partial \rho}\;\hat{z}-\frac{\partial A_{\varphi}}{\partial z}\;\hat{\rho}\Bigg) \\
\mathbf{\nabla \times W}=\mathbf{B_{W}}= -\Bigg(\frac{\partial W_{z}}{\partial \rho}-\frac{\partial W_{\rho}}{\partial z}\Bigg)\;\hat{\varphi}
\end{eqnarray*}

The {\bf field equations} follow:
\begin{eqnarray*}
\partial_{\rho}^{2}F+\partial_{z}^{2}F+\frac{\partial_{\rho}F}{\rho}-\frac{g_{1}}{2}\Big(F^{2}-1\Big)F-\frac{g_{3}}{2}P^{2}F-{}
 \nonumber\\
{}\Bigg[ \Bigg(qW_{\rho}-M\frac{zcos^{2}\Theta}{(\rho-a)^{2}}\Bigg)^{2}+\Bigg(qW_{z}+M\frac{cos^{2}\Theta}{(\rho-a)}\Bigg)^{2}\Bigg]F=0 \\
\partial_{\rho}^{2}P+\partial_{z}^{2}P+\frac{\partial_{\rho}P}{\rho}-\Big(eA_{\varphi}+N\Big)^{2}\frac{P}{\rho^{2}}-
\frac{g_{2}}{2}\Big(P^{2}-u^{2}\Big)P-\frac{g_{3}}{2}F^{2}P=0 \\
\partial_{\rho}^{2}A_{\varphi}+\partial_{z}^{2}A_{\varphi}-\frac{\partial_{\rho}A_{\varphi}}{\rho}
-2eP^{2}\Big(eA_{\varphi}+N\Big)=0 \\
\partial_{z}^{2}W_{\rho}-\partial_{z}\partial_{\rho}W_{z}-2qF^{2}\Bigg(qW_{\rho}-M\frac{zcos^{2}\Theta}{(\rho-a)^{2}}\Bigg)=0 \\
\partial_{\rho}^{2}W_{z}+\frac{1}{\rho}\Big(\partial_{\rho}W_{z}-\partial_{z}W_{\rho}\Big)-\partial_{\rho}\partial_{z}W_{\rho}
-2qF^{2}\Bigg(qW_{z}+M\frac{cos^{2}\Theta}{(\rho-a)}\Bigg)=0
\end{eqnarray*}

We can also write down the {\bf currents} associated with $\phi$ field, namely $j_{\rho}^{\phi}$ and $j_{z}^{\phi}$ and the total current $\mathcal{I}^{\phi}$
out of these as well as the supercurrent $\mathcal{I}^{\psi}$ associated with the $\psi$ field. These are
\begin{equation}
\mathcal{I}^{\phi}= \sqrt{(j_{\rho}^{\phi})^{2}+(j_{z}^{\phi})^{2}}\; ,\;\;\; \mathcal{I}^{\psi}= -\frac{2eP^{2}}{\rho}(eA_{\varphi}+N)
\end{equation}
where
\begin{equation}
j_{\rho}^{\phi}= -2qF^{2}(qW_{\rho}+M\partial_{\rho} \Theta), \;\;\;  j_{z}^{\phi}= -2qF^{2}(qW_{z}+M\partial_{z} \Theta)
\end{equation}

\begin{figure}
\centering
\includegraphics[scale=0.45 ,angle=270]{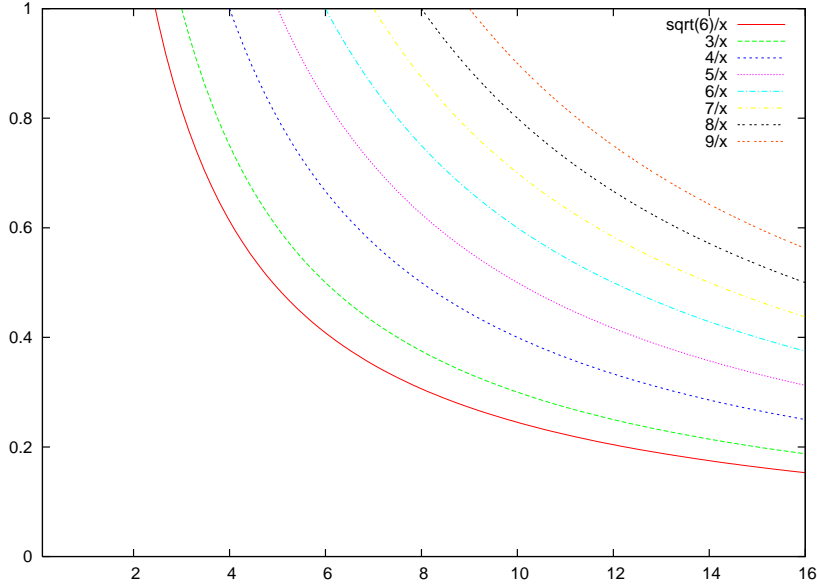}
\caption{\small  For different values of $\sqrt{g_{1}}$, we plot the lower bound over which the condition (\ref{cd}) is satisfied.
The plot is $P_{max}$ vs. $e$ and we plot the function $P_{max}=\frac{\sqrt{g_{1}}}{e}$.\label{param} }
\end{figure}
As explained in the introduction, Meissner effect is of crucial importance for the stability of the torus-like object. The magnetic fields
produced  by the supercurrent $\mathcal{I}^{\psi}$ penetrate into the toroidal defect in a distance dictated by the penetration depth.
In this theory, the $U(1)_{\mathbf{A}}$ symmetry breaks inside the string and the photon acquires mass $m_{\mathbf{A}}^{2}=e^{2}<P>^{2}$
where $<P>$ is the expectation value of the charge condensate in the vicinity of the string core. There is the superconducting sector of the defect.
The penetration depth is $\lambda=\frac{1}{m_{\mathbf{A}}}=\frac{1}{e<P>}$. But $<P>=P_{max}\leq u$ thus, one can have an estimate for $\lambda$, with
a lower bound for its value. This is $\lambda \geq \frac{1}{eu}$ where the equality holds when $P_{max}=u$.
On the other hand, one can also compute an upper bound for the thickness of the defect. We know that $r_{\phi}=\frac{1}{m_{\phi}}=\frac{1}{\sqrt{g_{1}}}$.
If our concern is to search for stable rings, a reasonable step is to demand the penetration depth to be smaller than the string thickness which means
\begin{equation}
\label{cd}
\lambda < r_{\phi} \Rightarrow \frac{1}{e P_{max}} < \frac{1}{\sqrt{g_{1}}} \Rightarrow e^{2} P_{max}^{2}> g_{1}
\end{equation}
This is the condition  needed in this case. From the above condition, we get the diagram shown in fig.\ref{param} where
one can see the area where it's more possible to find stable solutions if they exist. 
The numerical results we  present  later, are what we found while searching inside this region.

A way to derive {\bf virial} relations is through Derrick's theorem.
The virial relation for the field $\phi$ of our model, must have the constraint $\phi_{\kappa}(\rho=a, z)=\phi (\rho=a, \kappa z) =0$.
Consider the rescalings $\rho\rightarrow \rho$, $z\rightarrow \kappa z$, 
$F_{\kappa}\rightarrow F$, $P_{\kappa}\rightarrow P$, $A_{\varphi_{\kappa}} \rightarrow A_{\varphi}$,
$W_{\rho,z_{\kappa}} \rightarrow \kappa W_{\rho,z}$. Then, we find the minimum through the relation $\frac{\partial E}{\partial \kappa} =0$ when $\kappa =1 $.
The virial relations we use in our model, in order to check our numerical results follow. If we define
\begin{eqnarray*}
I_{1}&=& 2\pi v_{1} \int_{0}^{\infty} \rho d\rho \int_{-\infty}^{\infty} dz \Bigg[\frac{1}{2}(\partial_{\rho} W_{z} - \partial_{z} W_{\rho})^{2}+
             \frac{1}{2\rho^{2}} (\partial_{z} A_{\varphi})^{2}+(\partial_{z} P)^{2}+(\partial_{z} F)^{2}+{}
          \nonumber\\
        {}&+&2F^{2}\Bigg( qW_{\rho}(qW_{\rho}+M\partial_{\rho}\Theta )+(qW_{z}+M\partial_{z} \Theta)^{2}\Bigg)
\Bigg] \\
I_{2}&=&  -2\pi v_{1} \int_{0}^{\infty} \rho d\rho \int_{-\infty}^{\infty} dz \Bigg[\frac{1}{2\rho^{2}}(\partial_{\rho}A_{\varphi})^{2}+(\partial_{\rho} F)^{2}+(\partial_{\rho} P)^{2}
              +\frac{P^{2}}{\rho^{2}}(eA_{\varphi}+N)^{2}+{}
        \nonumber\\
{}&+&(\partial_{z}W_{\rho})(\partial_{\rho}W_{z}- \partial_{z}W_{\rho})+F^{2}\Bigg( (qW_{\rho}+M\partial_{\rho} \Theta)^{2} + (qW_{z}+M\partial_{z} \Theta)^{2}\Bigg)+{}
	                                                                                 \nonumber\\
     {}&+&\frac{g_{1}}{4}(F^{2}-1)^{2} + \frac{g_{2}}{4} (P^{2} -u^{2})^{2} +\frac{g_{3}}{2} F^{2}P^{2}-\frac{g_{2}}{4}u^{4}\Bigg] 
 \end{eqnarray*}	
we must have $I_{1}+I_{2}=0$. We define the index $V=\frac{||I_{1}|-|I_{2}||}{|I_{1}|+|I_{2}|}$ and we want its value to be as small as possible.
We can derive many other virial relations by assuming generally for a field $\phi$, the ``double'' rescaling $\phi (\vec{x}) \rightarrow  \kappa \phi(\mu \vec{x})$
and then demanding $\frac{\partial E}{\partial \kappa}|_{\kappa = 1 = \mu} =0 =\frac{\partial E}{\partial \mu}|_{\kappa = 1 = \mu}$. For example,
consider the following rescalings $\rho\rightarrow \rho $, $z\rightarrow \mu z$, $F_{\kappa}\rightarrow F$, $P_{\kappa}\rightarrow \kappa P$,
$A_{\varphi_{\kappa}} \rightarrow A_{\varphi}$, $W_{\rho ,z_{\kappa}} \rightarrow  W_{\rho,z}$. 
We have
\begin{eqnarray*}
I_{3}&=& 2\pi v_{1} \int_{0}^{\infty} \rho d\rho \int_{-\infty}^{\infty} dz \Bigg[ \frac{1}{2\rho^{2}} (\partial_{z} A_{\varphi})^{2}+(\partial_{z} P)^{2}+
               (\partial_{z} F)^{2}+{}
			                                          \nonumber\\
     {}&+&2F^{2}\Bigg( M\partial_{z} \Theta (qW_{z} +M\partial_{z} \Theta )\Bigg)+\frac{1}{2}(\partial_{\rho}W_{z}-\partial_{z}W_{\rho})^{2}\Bigg]  \\
I_{4}&=& -2\pi v_{1} \int_{0}^{\infty} \rho d\rho \int_{-\infty}^{\infty} dz \Bigg[\frac{1}{2\rho^{2}} (\partial_{\rho} A_{\varphi})^{2}+(\partial_{\rho} P)^{2}
                                   +(\partial_{\rho} F)^{2}+\frac{P^{2}}{\rho^{2}}(eA_{\varphi} +N)^{2}+{}
	                                                                   \nonumber\\
        {}&+&\partial_{\rho}W_{z}(\partial_{\rho}W_{z}-\partial_{z}W_{\rho})+F^{2}\Bigg( (qW_{\rho}+M\partial_{\rho}\Theta)^{2}+(qW_{z}+M\partial_{z}\Theta)^{2}\Bigg) +{}
		\nonumber\\
        {}&+&\frac{g_{1}}{4}(F^{2}-1)^{2} + \frac{g_{2}}{4} (P^{2} -u^{2})^{2} +\frac{g_{3}}{2} F^{2}P^{2}-\frac{g_{2}}{4}u^{4}\Bigg] 
\end{eqnarray*}
together with
\begin{eqnarray*}
I_{5}&=& 2\pi v_{1} \int_{0}^{\infty} \rho d\rho \int_{-\infty}^{\infty} dz \Bigg[ 2(\partial_{\rho} P)^{2}+ 2(\partial_{z} P)^{2}
                  +\frac{2P^{2}}{\rho^{2}} (e A_{\varphi} +N)^{2} \Bigg]  \\
I_{6}&=& 2\pi v_{1} \int_{0}^{\infty} \rho d\rho \int_{-\infty}^{\infty} dz \Bigg[g_{2}\Big( P^{2}-u^{2}\Big) P^{2} + g_{3}F^{2}P^{2}\Bigg]
\end{eqnarray*}
where, as above, we must have $I_{3}+I_{4}=0=I_{5}+I_{6}$.

\subsection {{\normalsize Numerical results}}
Energy minimization algorithm is used to minimize the energy functional of (\ref{funcu1}). The algorithm is 
written in C. One can find details about the algorithm on page 425 of \cite{c6} but, briefly, the basic idea is this:
Given an appropriate initial guess, there are several corrections to it, having as a criterion the minimization of the energy
in every step. When the corrections on the value of the energy are smaller than $\sim 10^{-8}$ the program stops and we get the final results.
A 90$\times$20 grid for every of the five functions is used, that is, 90 points on $\rho$-axis and 20 on $z$.
Here, we begin with fixed torus radius $a$. Then, the configuration with minimum energy for this $a$ is found. Other values of $a$
are chosen as well and the same process goes on until we plot the energy vs. the torus radius $E(a)$. It would be very
interesting to find a non-trivial minimum of the energy (on $a_{min}\neq 0$), which would correspond to stable toroidal defects with radius $a_{min}$.

\begin{figure}[t]
\centering
\includegraphics[scale=0.45]{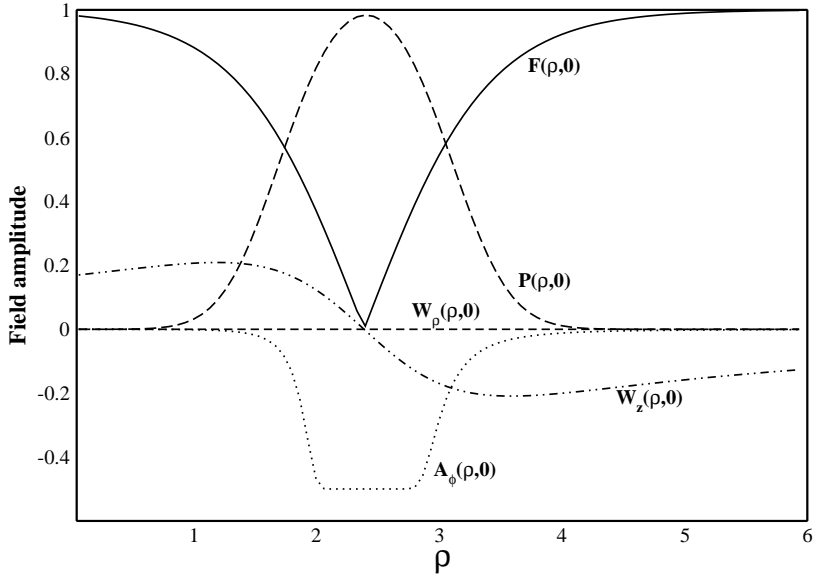}
\caption{\small  A typical plot of the initial guess we use for the five fields for the lowest winding state $M=1, N=1$
on the $z=0$ plane. \label{ig}}
\end{figure}

The {\bf initial guess} (fig.\ref{ig}) we use for our computation is:
\begin{eqnarray*}
F(\rho,z)&=& \tanh((\rho-a)^{2}+z^{2})^{M/2} \\
P(\rho,z)&=& \tanh(\rho^{N}) (1-\tanh((\rho -a)^{2}+z^{2}) \\
A_{\varphi}(\rho,z)&=& -\frac{N}{e}\tanh\Bigg(\frac{\xi \rho^{2}}{((\rho-a)^{2}+z^{2})^{2}}\Bigg) \\
W_{\rho}(\rho,z)&=&\frac{Mz\cos^{2}\Theta}{q(\rho-a)^{2}}\Bigg(\frac{(\rho-a)^{2}+z^{2}}{(\rho-a)^{2}+z^{2}+(a^{2}/4)}\Bigg)^{2} \\
W_{z}(\rho,z)&=&-\frac{M\cos^{2}\Theta}{q(\rho-a)}\Bigg(\frac{(\rho-a)^{2}+z^{2}}{(\rho-a)^{2}+z^{2}+(a^{2}/4)}\Bigg)
\end{eqnarray*}
where $\xi$ a constant.
This initial guess also satisfies the appropriate asymptotics
\begin{itemize}
\item{near $\rho =0$: 
\begin{equation}
F\neq 0 , \;\; P\sim \rho^{N}, \;\; A_{\varphi} \sim \rho^{2}f(z)
\end{equation} }
\item{near $(\rho =a , z=0)$: 
\begin{equation}
F\sim \tilde{\rho}^{M/2}, \;\; W_{\rho}=0=W_{z}
\end{equation} }
\item{at infinity: 
\begin{eqnarray}
F&\sim& 1-\mathcal{O}(e^{-\sqrt{\tilde{\rho}}}),\;\;\; P\sim \mathcal{O}(e^{-\sqrt{\rho^{2} +z^{2}}}) {}
  \nonumber\\
{}W_{\rho} &\sim& -\frac{M}{q}\partial_{\rho}\Theta|_{\infty} +\mathcal{O}(e^{-\sqrt{\tilde{\rho}}}),\;\;\; 
W_{z} \sim -\frac{M}{q}\partial_{z}\Theta|_{\infty} +\mathcal{O}(e^{-\sqrt{\tilde{\rho}}})
\end{eqnarray} }
\end{itemize}
where $\tilde{\rho}\equiv (\rho -a)^{2}+z^{2}$.

\begin{figure}
\centering
\includegraphics[scale=0.45]{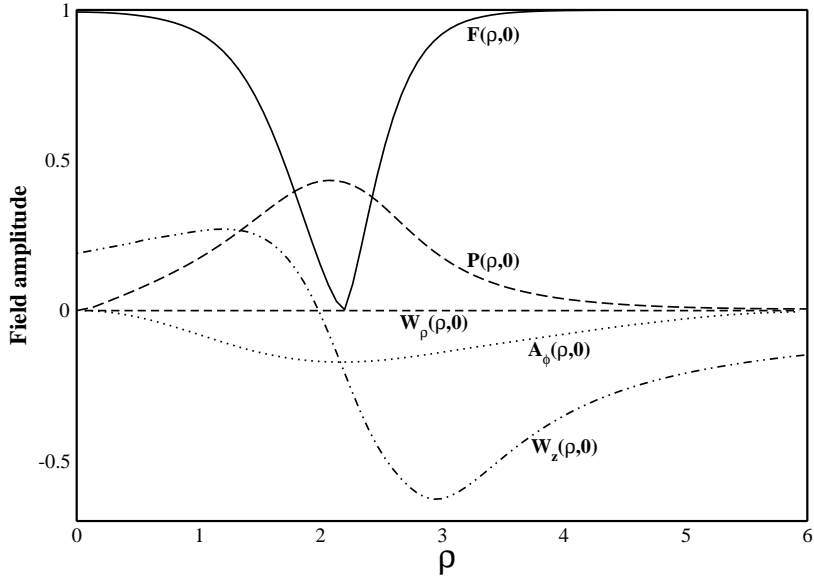}
\caption{\small  Typical plot of the final configuration of fields on the $z=0$ plane. Parameters are
$M=1$, $N=1$, $e=5$, $q=m_{W}=2$, $g_{1}=14$, $m_{\phi}=\sqrt{g_{1}}=3.74$, $g_{2}=12$, $g_{3}=14$, $u=1$, 
$m_{\psi}= \sqrt{(g_{3}-g_{2}u^{2})/2}= 1$, $v_{1}=7.5\cdot 10^{-3}$. Energy $E=0.72$ and virial is $1.5\cdot 10^{-3}$.  \label{p1}}
\end{figure}

For fixed torus radius (i.e. here $a\approx 2.2$) we present a typical graph of the final configuration of the
lowest energy (see fig.\ref{p1}). 
We also present the plot of the energy of the system vs. the radius
of the toroidal object which reveals the instability of the system as there is no non-trivial minimum (see fig.\ref{enercond}).

From the results, we can point out a few things. Firstly, the greater the value of $e$ we use, the stronger the supercurrent becomes.
Secondly,  the greater the value of $e$ we use, the lower the radius $a$ where the supercurrent quenches (fig.\ref{jepe}). 
These are expected as the increase of $e$ makes the condition of eq.\ref{cd}
stronger, something which means that the mass of the photon increases and the penetration depth
decreases at the same time. It is also reasonable that a stronger current can ``defend''  the defect, against the magnetic field,
a little longer. 
Another observation is that as the difference $g_{1}-g_{2}$ decreases, that is to say, as $g_{2}$ increases and becomes close to $g_{1}$,
we see that the maximum current increases and the quenching takes place again at lower $a$ (fig.\ref{jepg2}). This is expected as one can see from the potential
(eq.\ref{potu1}) of the energy functional, because the stronger the coupling $g_{2}$ is, the more important the relevant term $g_{2}(P^{2}-u^{2})^{2}/4$ becomes.
The latter has as a consequence, the increase of $P_{max}$  which counteracts the effects from the increasing $g_{2}$ coupling.

The parameter space where we searched, starts from $g_{1} =4 $ ($m_{\phi} =2$). 
In order to search the model, we reached values around $g_{1} = 30$ ($m_{\phi}\approx 5.48$)
over which, $e$ has to be very large in order to respect the condition (\ref{cd}). There is also the
fact that great values of $e$ in general is an unwanted feature since we use semiclassical approach.
As it concerns the other couplings we have $g_{3}= g_{1}$ and  $g_{2}=g_{1}-k$ with $0.5\leq k \leq 8$, ($u=1$).

\section {{ \normalsize Explanation concerning the instability of the vortex ring}}
From the  condition  in eq.\ref{cd}, it is understood that we are enforced to  lower $g_{1}$ as much as possible and/or  increase $e$. But
these steps are not as easy as they might look. There are some limits.
The lower bound on the value of $g_{1}$ has
a reasonable explanation. For low values of $g_{1}$, the coupling $g_{2}$ is also low (because $g_{1}>g_{2}u^{4}$).
Now,  when $g_{2}$ is small enough, the changes on the term $g_{2}(P^{2}-u^{2})^{2}/4$ are unimportant for the energy, in comparison
to the term $(\partial_{\rho}P)^{2}$. In that case, the lowering of the last term minimizes the energy, something which means that
$P \rightarrow 0$. Indeed, this is computationally observed.
There is also a lower limit on $e$ which is reasonable because the lowering of $e$ results to an increasing
penetration depth.

\begin{figure}[t]
\centering
\includegraphics[scale=0.45]{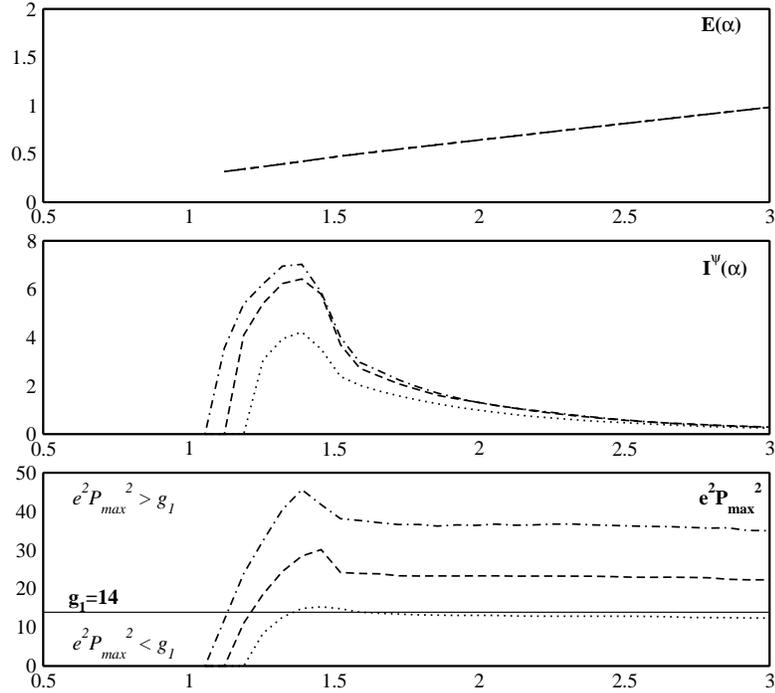}
\caption{\small  The top graph is the energy vs. the torus radius $E(a)$. The middle graph
is the supercurrent $\mathcal{I}^{\psi}$ vs. $a$. On that graph one can clearly see current quenching. The bottom graph is the
quantity $e^{2}P_{max}^{2}$ vs. $a$ (or $m_{\mathbf{A}}$ vs. $a$) where one can see the area in which the condition of eq.\ref{cd} holds
(lines over the $g_{1}$-limit line). The resistance from the magnetic field can be seen as a  sharp increase on the supercurrent.
Dotted lines are for $e=6$, dashed for $e=8$ while dashed and dotted for $e=10$. All plots are for $(M,N,u,g_{1},g_{2},g_{3})$=$(1,1,1,14,12.5,14)$.\label{jepe}}
\end{figure}

\begin{figure}[t]
\centering
\includegraphics[scale=0.45]{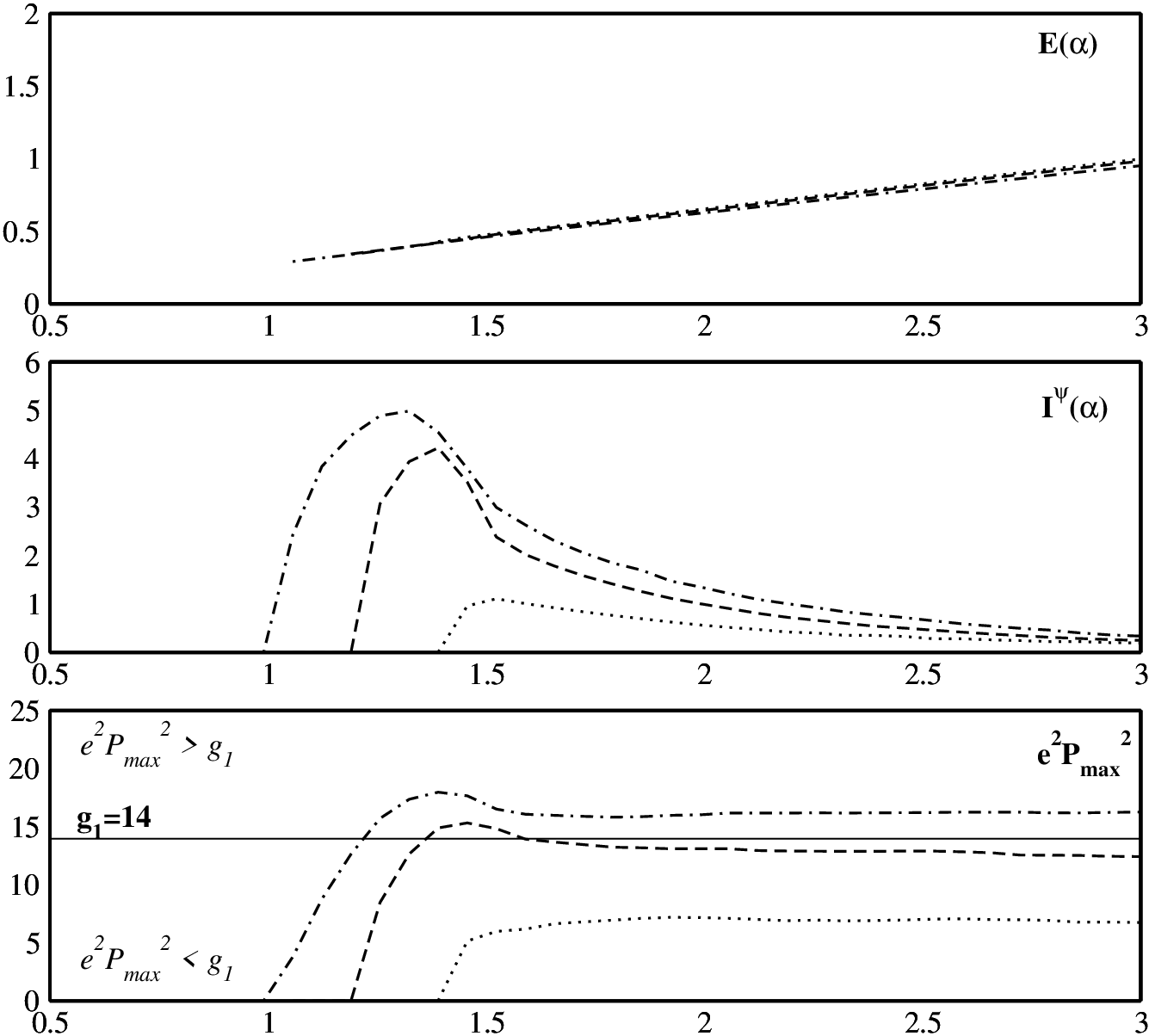}
\caption{\small  The top graph is the energy vs. the torus radius $E(a)$. The middle graph
is the supercurrent $\mathcal{I}^{\psi}$ vs. $a$. On that graph one can clearly see current quenching. The bottom graph is the
quantity $e^{2}P_{max}^{2}$ vs. $a$ (or $m_{\mathbf{A}}$ vs. $a$) where one can see the area in which the condition of eq.\ref{cd}  holds
(lines over the $g_{1}$-limit line). The resistance from the magnetic field can be seen as a  sharp increase on the supercurrent. 
Dotted lines are for $g_{2}=12$, dashed for $g_{2}=12.5$ while dashed and dotted for $g_{2}=13$. All plots are for $(M,N,u,e,g_{1},g_{3})$=$(1,1,1,6,14,14)$.\label{jepg2}}
\end{figure}

\begin{figure}[t]
\centering
\includegraphics[scale=0.45]{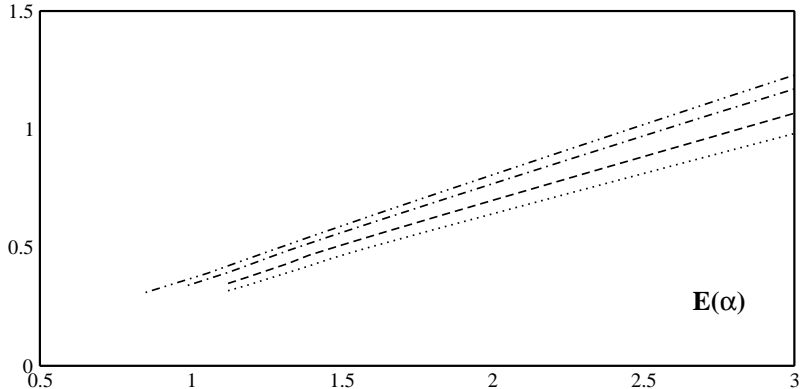}
\caption{\small  The plot of the energy of the system vs. the radius of the torus for four
different sets of parameters. Dotted is for $(g_{1}, g_{2}, g_{3})$=$(14,12.5,14)$, dashed is for $(18,15,18)$, dashed and dotted is for $(25,20,25)$
while dashed with two dots is for $(30,24,30)$. For all data sets we have $(e,M,N,u)$=$(10,1,1,1)$.
As one can observe, there exists no minimum.\label{enercond}}
\end{figure}

We searched for stable objects for values over these  limits described above. The computational results are exhibited
in figs.\ref{jepe}-\ref{enercond} and as we see, these objects are unstable. We argue that the explanation for the instability is
{\bf current quenching} and that, only for high values of $e$. For  values of $e$,  of the order of $1$, we have,
according to eq.\ref{cd}, that the penetration depth is much bigger than the string thickness thus, stability is out of the question.
The latter is also computationally observed.

Now, we base our aspect about quenching on qualitative as well as quantitative arguments. 
We observe that as the torus shrinks, the supercurrent suddenly drops to zero which signals
the destruction of the defect. Just before the sharp drop, we notice that the supercurrent rises with increasing rate. This must be due to the resistance the torus meets
from the magnetic lines as it shrinks. 
One can observe that as the supercurrent increases and the condition of eq.\ref{cd} is satisfied at the same time (i.e. see dashed and dotted line of fig.\ref{jepe}),
suddendly the current is lost. This can be explained only through current quenching.
The above phenomenon is not observed when eq.\ref{cd} is not satisfied (i.e. see dotted line of
fig.\ref{jepg2}). There, as the magnetic lines penetrate the ring, they meet almost no resistance since $\lambda$ is much greater than $r_{\phi}$.
Another observation which supports our quenching argument is that, rough estimations on the maximum current a string can sustain, lead us to the
following formula (see p.129 of \cite{c8}) which makes a small overestimation in order to have an upper limit:
\begin{equation}
\label{maxj}
\mathcal{I}^{\psi}_{max} < \sqrt{\sigma} e u
\end{equation}

where $\sigma \equiv \int\int d\rho dz\; P^{2}$. In figs.\ref{jepe}-\ref{jepg2}, the maximum value of the supercurrent is close to the limit
of the estimation of eq.\ref{maxj}. The table below gathers the estimated $\mathcal{I}^{\psi}_{est.}$ (according to eq.\ref{maxj})
and the computed maximum supercurrent ($\mathcal{I}^{\psi}_{com.}$) for the parameters of these figures.
\newline
\newline
\begin{tabular}{|l|l|l|l||l||l||}
\hline
$e$ & $g_{1}$ & $g_{2}$ & $g_{3}$ &  $\mathcal{I}^{\psi}_{est.}$ & $\mathcal{I}^{\psi}_{com.}$ \\
\hline
\hline
6 & 14 & 12.5  & 14 & 4.6  & 4.3    \\
\hline
6 & 14 & 13    & 14 & 5.2  & 5.0    \\
\hline
8 & 14 & 12.5  & 14 & 6.7  & 6.4    \\
\hline
10 & 14 & 12.5 & 14 & 8.0  & 7.2    \\
\hline
\end{tabular}
\newline
\newline

Finally, one can make an estimation of the value of the supercurrent which would stabilize the ring, namely $\mathcal{I}^{\psi}_{stab.}$.
This can be done as follows. As explained in the introduction, there is the tension of the string which shrinks the loop and
the magnetic field which opposes this tendency. When  the ring is stabilized we have $E_{tension}=E_{magnetic}$. Here, $E_{tension}\sim 1$
and $E_{magnetic}=2\pi v_{1}\int \rho d\rho \int dz (B_{A}^{2}/2)$. Without any calculation, one can point out that, since
the {\em total} energy in the quenching radius is below $E_{tension}=1$, then the $E_{magnetic}$ which is a fraction of it, would
be even smaller. Recall that $B_{A}\propto \mathcal{I}^{\psi}$, which means that we need a $\mathcal{I}^{\psi}_{stab.}$ which is well above
the maximum current we can have inside the defect. Calculations of the magnetic energy are in agreement with the above observation and
place its value around $E_{magnetic}\sim 5\cdot 10^{-3} << E_{tension} \sim 1$. This translates to the following conclusion:
$\mathcal{I}^{\psi}_{stab.} \gtrsim 10\cdot \mathcal{I}^{\psi}_{est.}$.

Thus, we find out that the current needed for stabilization, is at least ten times bigger than the value of the maximum current our
string can sustain. We also observe that our computational maximum current values are close to the theoretical estimations about quenching.
This means, that we will have current quenching as an ``obstacle'' towards stabilization, since the supercurrent will not be enough
in order to create the magnetic field needed.

\section {{\normalsize Conclusions}}
Future experiments on LHC could answer whether metastable particle-like solitons exist in MSSM or 2HSM {\em or} not. 
On \cite{c3} there is a search for spherically symmetric solitons in the frame of the 2HSM but with a simplified potential.
Here we search for axially symmetric solitons which, if stable, will have a mass of few TeV \cite{c2}.
We considered the $U(1)_{A}\times U(1)_{W}$ model, where the existence of the vortex is ensured, 
for topological reasons. There, we searched for stable toroidal strings. We present and analyse our observations.
This paper tries to answer to expectations having to do with observations of stable axially symmetric solitons which would be possible to detect in later experiments 
of LHC. For relatively small values of $e$ ($\sim 1$), the system
seems to have  no stable vortex rings. In fact, this instability is present even in other parameter areas where we searched (i.e. $e\geq 6$ see figs.\ref{jepe},\ref{jepg2}).
We explain why we believe that the main reason of instability is current quenching.

\section {{\normalsize Acknowledgments}}
The author is very thankful to Professor T.N.Tomaras for fruitful conversations and useful advice. Work supported in part
by the ``Superstrings'' Network with EU contract MRTN-CT-2004-512194.

\end{document}